
\magnification=1200
\rightline {RU--91--50}
\rightline {Novemeber 26, 1991}
\vskip 1.5cm

\centerline{\bf THREE DIMENSIONAL PERIODIC $U(1)$ GAUGE THEORY AND
STRINGS\footnote{*}{\rm Talk delivered at the International Symposium on
Lattice Field Theory, Latt91; National Laboratory for High Energy Physics,
Tsukuba, Japan; Nov. 5-10, 1991.}}
\vskip 1cm
\centerline {by}
\vskip 1cm
\centerline{Herbert Neuberger}
\vskip .2cm
\centerline{\it Department of Physics and Astronomy}
\centerline{\it Rutgers University}
\centerline{\it Piscataway, NJ 08855-0849, U.S.A.}
\vskip 2cm
\centerline {\bf Abstract}
\vskip 1cm
It will be argued that among the known systems in three
dimensions that have string like excitations periodic U(1) pure gauge theories
are the most likely candidates to lead to a string representation of their
universal properties. Some recent work with F. David will also be reviewed.
\vskip .3cm
\vfill
\eject
Three dimensional periodic $U(1)$ gauge theories confine due to the dual
Meissner effect [1,2]. Lattice formulations as well as continuum formulations
lead to very similar physics. But this similarity has never been made very
precise. On the basis of semiclassical calculations the underlying dynamics is
seen to be connected to the three dimensional Sine-Gordon model. Normally, one
would view
such a model as a representative of the universality class in three dimensions
of all local models that have a single component scalar field and a global
symmetry group $Z$. However, no
special fixed point is known in this class -- it seems that in the infrared the
global symmetry either gets elevated to $R$ (with some fine tuning) or,
generically, disappears completely.

If we accept that indeed no fixed point with a symmetry strictly $Z$ exists in
3d, we conclude that the above similarity of confinement mechanisms cannot be
made precise in any ordinary field theoretical continuum limit by the usual
mechanism of Renormalization Group universality. Of course, one can simply say
that there is nothing strictly universal about the 3d dual Meissner confining
phase, period. In this talk I shall explore the opposite point of view, namely
that there indeed is something universal, and only the correct framework for
abstracting these universal features hasn't been found yet.

Any confining 3d $pU(1)$ (three dimensional periodic $U(1)$) has string-like
excitations. By this I mean quasi-stable states whose wave-functional shows a
concentration of energy density along a relatively smooth, one dimensional
curve.

The suggestion I would like to make is that maybe, the missing abstraction of
the universal features of the various versions of the Meissner confinement
mechanisms in the usual field theoretical framework can be found in a string
theory. Such a theory would have, as fundamental objects, ``bare'' excitations
associated with mathematical smooth closed curves, which interact during their
evolution by spanning two dimensional surfaces of higher genus embedded in 3d
Euclidean or Minkowski flat space.

There is little doubt that such surfaces cannot ``go through each other''
without an effect on the Feynman probability amplitude. In ordinary field
theory the paths described by the point like excitations are transparent when
they cross each other and this is crucial for having a local field theory
associated in a precise way to these paths.

But, there is a possibility that the essential features of the effects of
surfaces going through each other, can be incorporated by adding more,
intrinsic, degrees of freedom to the string. What looks as interaction at
surface crossings when the intrinsic degrees of freedom have been averaged out
or frozen may appear as simple statistics when the intrinsic degrees of freedom
are kept. For example, as emphasized recently by Polyakov, this is how the 2d
Ising model reappears as a free field theory and most of this can be
generalized to the 3d Ising model. To do this in a precise way is the crux of
the matter when one attempts to associate a string theory with the strong
coupling expansion surfaces appearing in any gauge model [3]. However, it may
be that this cannot be done precisely at all on the lattice, but, nevertheless,
the few ``glitches'' that one gets are irrelevant in the continuum limit and
the correspondence to a string theory indeed does hold in the continuum. For
ordinary four-dimensional gauge theories the prospects for this to actually
work in a direct way are pretty dim because the field theory has unmistakable
point-particle-like behavior in the ultra-violet and this seems to be
impossible to reproduce in a theory that is ``stringy'' all the way down to
zero distance. If there exists a correspondence to a string theory it must be
of a less direct kind [4]. A more direct correspondence might hold for N=4
supersymmetric Yang Mills: its UV finiteness may permit tuning a coupling to a
stringy ``double scaling limit'' for its Feynman diagrams organized in powers
of $1/N_c$ with $N_c$ given by the gauge group.

Nevertheless, Polyakov and others went ahead and suggested that the broken
phase side of the second order transition in the 3d Ising ferromagnet is
describable by a string theory and some of the nontrivial critical exponents of
the 3d Ising model can be calculated within this string theory [5]. This would
be a very beautiful theoretical development; however, unlike in the case of
$pU(1)$, the universal features of the transition are in this case describable
by an ordinary (albeit strongly interacting) field theory, and therefore the
string is not necessarily needed to abstract the generic features of the
transition.

3d $pU(1)$ and 3d Ising are not that different; the main dissimilitude is that
while Ising strings are unoriented $pU(1)$ strings do carry an orientation. One
may guess that $pU(1)$ strings are represented by some kind of a complexified
version of the fermionic representation of Ising strings. I am unaware of any
specific attempt to derive such a representation (within some specific lattice
model, for example).

If either 3d Ising or 3d $pU(1)$ gauge theories have some limit where a
complete set of observables can be extracted and represented by a string theory
one has to ask what the coupling constant of this string theory would be.
Without knowing exactly what the elementary excitations along the string are we
cannot realistically hope to answer this question. Indeed, if we try to
identify the surfaces appearing in the strong coupling expansions of the gauge
versions of either model with world histories of strings (an identification
that has a correct analogue in 2d), we very soon are faced with the occurrence
of self-crossings and singular lines which render ambiguous the genus to be
associated with the so inflicted surfaces [3].

Let us adopt the following strategy then (in our search to identify a ``bare''
string coupling): As a first step find a model where all singular lines
disappear; this can be done at the price of absolute repulsion, i.e.
self-avoidance. It is an open and important question whether this
self-avoidance can be replaced by some interaction of purely statistical
origin. If this can be done, it is likely that the genus of the surfaces in the
new theory will be, with probability one, equal to the sum over genera of the
connected components of the self-avoiding surface. Having identified the genus
of any strong coupling diagram it is possible then to weigh the diagrams by a
fixed factor raised to the power of the genus. This factor plays the role of a
``bare'' (i.e. defined at the cutoff scale) string coupling constant. Simple
arguments show that, in the Ising case this amounts to the addition of local
multispin interactions to the basic Ising action [6]. By field theory
universality these terms, if sufficiently weak, have no effect in the continuum
limit. Hence, if there exists a string representation of the broken phase of
continuum $\phi^4$ in 3d it has no freely adjustable string coupling constant,
and in that case is very unlikely to be a weakly coupled string, unless, for
some special reason, the string coupling must actually vanish in which case we
end up with a free string theory.

It was pointed out by David [7] that formulating the Ising model on the f.c.c.
lattice produces exactly the absolute self-avoidance effect that we required
above; moreover, David showed that the bare string coupling, if equal to
${1\over{N}}$ with integer $N$, could be exactly incorporated into a $O(N)$
lattice gauge theory with a single ``plaquette'' action of a special form
defined on the dual to the f.c.c. lattice, a lattice whose elementary cell is a
rhombic dodecahedron (also the shape of the first Brillouin zone of a body
centered cubic lattice). This ``rhombic dodecahedral lattice'' is not a Bravais
lattice.

In view of the above we would like to investigate whether an analogous
formulation can be found for the $pU(1)$ case [8]. Again self-avoidance can be
ensured by going to a particular gauge theory defined on a rhombic dodecahedral
lattice with $U(N),~N \ge 4$, gauge group. ${1\over{N}}$ plays the role of a
``bare'' string coupling constant as before. The theory is dual to a $Z$ spin
model on the f.c.c. lattice with nearest neighbor interactions and with a
restriction limiting the possible differences between neighboring spins to
$0,~\pm 1$. It is hoped, but not at all established convincingly, that this
model is in the same ``pseudo--universality'' class as the 3d $Z$-ferromagnet,
or S.0.S., model. Again ${1\over {N}}$ appears in the dual model via the
coupling in front of a multi-spin local interaction. It is again plausible (but
in the absence of field theoretical universality less compellingly true) that,
if not too large,  $N$ has no measurable effect on the relevant long distance
physics and therefore, again, the physical coupling in the alleged string
theory would be fixed. Once more we end up with the choice between a free
string theory (due to some yet undiscovered symmetry principle) or a strongly
interacting one.

On the technical level the $U(N)$ model is amusing because it admits
Eguchi-Kawai reduction to an eight matrix model and needs no quenching. Thus,
at $N=\infty$, several things can be exactly calculated. In particular the
planar contribution to the free energy can be shown to vanish. The question
whether the model before reduction has phase transitions when $N$ increases has
not been investigated to date.

Both the $O(N)$ and the $U(N)$ gauge theories can be shown to be insensitive to
the addition to the action of terms that break local gauge invariance down to
the centers of the respective gauge groups. So here confinement is explicitly
solely center dependent. Moreover, what seems to matter about the center is
only whether it is $Z_2$ or $Z_k$ with an arbitrary $k\ge 4$. This is again in
agreement with standard lore.

I believe that the ``RSOS'' model David and I introduced [8] warrants a more
detailed examination and may provide us with some clues on whether the
similarities between the confinement mechanisms in various realizations of
$pU(1)$ gauge theories in 3d can be abstracted into something universal within
the framework of string theory. Our present knowledge, in particular the
rigorous and beautiful work of G\" opfert and Mack [9] indicate that this
cannot be done by taking an ordinary, field theoretical, continuum limit.

As a first step towards testing the above speculations one might try to use the
semiclassical methods pioneered by Polyakov and investigate higher spin
excitations in the instanton gas. If there is something exactly universal about
pure $pU(1)$ in 3d it most likely involves higher spin states.

\vskip 1cm

\noindent{\bf Acknowledgements.} This work has been supported in part by the
U.S. Department of Energy through Contract No. DE-FG05-90ER40559. I am grateful
to Fran\c cois David for a fruitful collaboration.

\vskip 1cm

\leftline{\bf References}
\vskip .5cm

\item {[1]} A. M. Polyakov, {\bf Gauge Fields and Strings}, Contemporary
Concepts in Physics, Volume 3, Harwood Academic Publisher, (1987).

\item {[2]} A. M. Polyakov, {\bf Nucl. Phys. B120} (1977) 429; M. E. Peskin,
{\bf Ann. Phys. (NY) 113} (1978) 122; R. Savit, {\bf Phys. Rev. Lett. 39}
(1977) 55; T. Banks, R. Myerson, J. Kogut, {\bf Nucl. Phys. B129} (1977) 493.

\item {[3]} I. Kostov, {\bf Nucl. Phys. B265[FS15]} (1986) 223; K. H. O'Brien,
J.-B. Zuber, {\bf Nucl. Phys. B253} (1985) 621.

\item {[4]} H. Neuberger, {\bf Nucl. Phys. B340} (1990) 703.

\item {[5]} A. M. Polyakov, {\bf Phys. Lett. 82B} (1979) 247, {\bf Phys. Lett.
103B} (1981) 211; Vl. S. Dotsenko, A. M. Polyakov, in {\bf Advanced Studies in
Pure Mathematics 16}, Academic Press (1988); Vl. S. Dotsenko, {\bf Nucl. Phys.
B285} (1987) 45; A. R. Kavalov, A. G. Sedrakayan {\bf Nucl. Phys. B285[FS19]}
(1987) 264; E. Fradkin, M. Srednicki, L. Susskind, {\bf Phys. Rev. D21} (1980)
2885; S. Samuel {\bf J. Math. Phys. 21} (1980) 2806.

\item {[6]} T. Hofs\" ass, H. Kleinert, {\bf J. Chem. Phys. 86} (1987) 3565.

\item {[7]} F. David, {\bf Europhys. Lett. 9} (1989) 575.

\item {[8]} F. David, H. Neuberger, {\bf Phys. Lett. B269} (1991) 134.

\item {[9]} M. G\" opfert, G. Mack, {\bf Comm. Math. Phys. 82} (1982) 545.

\vfill
\eject
\bye